\documentclass[prl,twocolumn,superscriptaddress,showpacs,showkeys]{revtex4}

\usepackage{amsfonts}
\usepackage{amsmath}
\usepackage{amssymb}
\usepackage{graphicx}
\usepackage{dcolumn}
\usepackage{bm}
\usepackage{hyperref}

\begin{document}
\title{Topological indices, defects and Majorana fermion in chiral superconductors}
\author{Daichi Asahi}
\affiliation{Department of Applied Physics, University of Tokyo, Tokyo 113-8656, Japan}
\author{Naoto Nagaosa}
\affiliation{Department of Applied Physics, University of Tokyo, Tokyo 113-8656, Japan}
\affiliation{Cross Correlated Materials Research Group (CMRG) and Correlated
Electron Research Group (CERG), ASI, RIKEN, Wako 351-0198, Japan}
\date{\today}

\begin{abstract}
We study theoretically the role of topological invariants to protect the
Majorana fermions in a model of two-dimensional chiral 
superconductors which belong to class D of topological periodic table. 
A rich phase diagram is revealed.
Each phase is characterized by the topological invariants for 2d ($Z$) and 1d ($Z_2$),
which lead to the Majorana fermion at the edge of the dislocation and the core of the vortex. 
Interference of the Majorana fermions originated from the different 
topological invariants is studied.  The stability of the Majorana fermion 
with respect to the interlayer coupling, i.e., in 3d, is also examined. 
\end{abstract}
\maketitle

Topological classification of the electronic states in solids
has shed a new light on the band structure of solids and also the
superconductivity~\cite{Review}.
Initiated by the proposal of $Z_2$ topological invariant 
and the quantum spin Hall effect in two-dimensional
spin-orbit coupled system~\cite{KaneMele} and its extension to 
three dimensions~\cite{Balents}, the more generic topological classification 
scheme, i.e., topological periodic table,
is now available based on the three symmetries, i.e.,
time-reversal ($\Theta$), particle-hole ($\Xi$), and 
chiral ($\Pi$) symmetries~\cite{Schnyder,KitaevPeriod}. 
Here $\Xi$ symmetry is
due to the superconductivity in the usual situation.  
There are 10 classes, and the topological invariant is specified 
depending on the dimensionality $d$ of the system
to characterize the nontrivial topological states.
Later, this ``10-fold way'' has been extended to 
the classification including the textures such as
domain walls and dislocations~\cite{Teo}. In this case, the 
dimensionality $D$ of the real space manifold surrounding 
the texture plays a key role, and $\delta=d-D$ is 
replacing $d$ for the topological classification~\cite{Teo}.
In the topological periodic table,  one can recognize the periodicity  
called Bott periodicity which relates the different classes in the ``diagonal'' direction. 
This periodicity can be understood by the continuous mapping of the 
Hamiltonian connecting the different classes and different $\delta$ by one. 
On the other hand, one can also consider the connection in the ``horizontal'' 
direction, i.e., dimensional reduction~\cite{Dim}. 
As an example, class AII has been characterized by $Z_2$ both
for  $\delta=3$ and $\delta=2$. The former corresponds to the 
$Z_2$ invariant corresponding to the strong topological insulator (TI) in 3d,
while the latter to the quantum spin Hall system in 2d ($D=0$) 
or the weak TI in 3d ($D=1$). 
Namely, a 3d system is characterized by
4 $Z_2$ topological invariants $\nu_0; \nu_1\nu_2 \nu_3$, where 
$\nu_0=1$ indicates the strong TI, while $\nu_0=0$ with at least one
nonzero $\nu_{1,2,3}$ means the weak TI.  These $\nu_{1,2,3}$ are the
topological invariants for $\delta=3-1=2$, and guarantee the
existence of gapless one-dimensional mode along the dislocation~\cite{Ran}.
  
This topological periodic table provides the powerful guiding 
principle also for the topological superconductors (TSs). 
Especially, the Majorana fermions expected to appear at 
the edge or the core of the vortex in TSs attract intensive 
interests from the viewpoint of the quantum information 
technology~\cite{Kitaev,Majorana1,Majorana2,Read,Ivanov,TSCreview}. 
Therefore, it is an important theoretical issue
to design the Majorana fermions in realistic systems.  
Proximity-induced superconductivity in 3d TI ~\cite{FuKane},
the superconductivity in a doped TI 
Cu$_x$Bi$_2$Se$_3$~\cite{Hor,FuBerg},
the possible TS in noncentrosymmetric systems with the Rashba spin 
splitting~\cite{Tanaka1,Fujimoto,Sau,QiHuges,Lutchyn,Alicea},
and $p$-wave superconductivity in Sr$_2$RuO$_4$~\cite{SRO} are the 
promising candidates as the host of the Majorana fermions.
As pointed out in ref.~\cite{AliceaReview}, most of the theoretical 
proposals for the Majorana fermions are based on the two models,
i.e., $p$-wave pairing in the one-dimensional spinless fermions 
(Kitaev model~\cite{Kitaev}) and the $p+ip$ pairing superconductor.   
 
In this paper, we study theoretically the topological invariants and 
their relation to the protected Majorana bound states in 
a model of class D chiral superconductors containing both the 
Kitaev model and $p+ip$ superconductor in the limiting cases.
The topological invariant for class D is
$0$ for $\delta=3$, $Z$ for $\delta=2$, and $Z_2$ for $\delta=1$.
Therefore, there is no "strong TS" in 3d, while
the 2d system is characterized by $Z$ topological invariant and
the 1d system by $Z_2$ topological invariant.
The purpose of the present Letter is to 
reveal the topological phase diagram characterized by these
invariants, and the associated Majorana fermions  
at the textures such as dislocations and vortices.  

We consider a generalized model of $p+ip$ wave superconductor 
on a square lattice in 2d. 
The Hamiltonian can be written as
$H = \sum_{\bm k} C^\dagger_{{\bm k}} H({\bm k}) C_{{\bm k}} $ with 
\begin{equation}
{\scriptstyle H({\bm k})}= \left(\begin{array}{cc}
{\scriptstyle 2t_{x}\cos k_{x}+2t_{y}\cos k_{y}-\mu } & {\scriptstyle d_{x}\sin k_{x}-id_{y}\sin k_{y} }\\
{\scriptstyle d_{x}\sin k_{x}+id_{y}\sin k_{y} }& {\scriptstyle \mu-2t_{x}\cos k_{x}-2t_{y}\cos k_{y}}
\end{array}\right),\label{eq:p+ip}
\end{equation}
and $C^\dagger_{\bf{k}} = (c^\dagger_{\bf{k}}, c_{- \bf{k}})$.
This $2\times 2$ Hamiltonian matrix can be expressed as
$H({\bm k}) =  H (k_{x}, k_{y}) = {\bm h}({\bm k}) \cdot {\bm \sigma}$
where ${\bm \sigma} = (\sigma^x,\sigma^y,\sigma^z)$ is the vector of Pauli matrices.
Since $(C^\dagger_{- \bf{k}} )^T = \sigma^x C_{\bf{k}}$, $H(\bf{k})$ should satisfy
\begin{equation}
H({\bm k}) = - \sigma^x H(-{\bm k})^T \sigma^x
\end{equation}
where $T$ means the transpose. 
This condition leads to the relation~\cite{AliceaReview}
\begin{equation}
h_{x,y}({\bm k}) = - h_{x,y}(-{\bm k}), \ \ \ \ h_z({\bm k}) = h_z(-{\bm k}).
\end{equation}
Therefore, for the time-reversal invariant momenta (TRIM),
which satisfy ${\bm k} \equiv - {\bm k}$, only $h_z({\bm k})$ cannot be
nonzero, i.e., ${\bm h}({\bm k})$ points either in $+z$ or $-z$ directions
as long as the gap opens, i.e., $| {\bm h}({\bm k})| > 0$.
There are 4 TRIMs in this 2d model, i.e., 
${\bm k}_\alpha = (0,0), (\pi,0), (\pi, \pi)$ and $(\pi,\pi)$, and the sign 
$s_\alpha = \pm 1$ of the corresponding $h_z$.  As will be discussed, 
$s_\alpha$ determines the $Z_2$ topological invariants and the parity of the
$Z$ invariant. 

Now let us start with the $Z_2$ invariant. For this purpose, 
let us consider the 1d Hamiltonian with fixed $k_{x}=\pi$ in eq.(\ref{eq:p+ip}), i.e.,  
\begin{equation}
{\scriptstyle H(k_{x}=\pi, k_{y}) }= \left(\begin{array}{cc}
{\scriptstyle -2t_{x}+2t_{y}\cos (k_{y})-\mu } & {\scriptstyle -id_{y}\sin (k_{y})}\\
{\scriptstyle id_{y}\sin (k_{y})}& {\scriptstyle \mu+2t_{x}-2t_{y}\cos (k_{y})}
\end{array}\right),
\end{equation}
which is nothing but the Kitaev model for a one dimensional 
topological superconductor~\cite{Kitaev}. 
The $Z_2$ topological invariant $\nu_x$ is related to the ``polarization''~\cite{Dim}.
\begin{equation}
\frac{\nu_{x}}{2}=P(k_x) =  \int^\pi_{-\pi} \frac{d k_y}{2\pi}  a_y (k_x,k_y)  \mod 1
\label{eq:pol}
\end{equation}
given by the Berry phase vector potential
$a_j (k_x, k_y) = (-i) \sum_{n:{\rm occupied}} \langle n {\bm k}| 
\partial/\partial {\bm k}_j | n {\bm k}\rangle$, and is given by 
$(-1)^{\nu_x} = s_{(\pi,0)} s_{(\pi,\pi)}$~\cite{AliceaReview,Turner}.
Therefore, we can easily obtain the $Z_2$ topological invariant $\nu_x$ as
\begin{equation}
\nu_{x}=\begin{cases}
1 & {\rm for}~~ \vert t_{x}+\frac{\mu}{2}\vert<\vert t_{y}\vert\\
0 & {\rm for}~~ \vert t_{x}+\frac{\mu}{2}\vert>\vert t_{y}\vert
\end{cases} \label{x-pi}
\end{equation}
The topological invariant $\nu_x'$ for $k_x=0$ can be also 
calculated in a similar way.
From these equations, it is clear that strengths of $d_{x}$ and $d_{y}$ are not 
related to the topological numbers if they have finite values. 
The $Z_2$ invariants $\nu_{y}, \nu_y'$ are obtained in the similar way.

On the other hand, the $Z$ topological invariant $\nu$ 
is nothing but the Chern number, i.e., the wrapping number of 
the mapping from the 1st Brillouine zone of ${\bm k}$ to the 
unit sphere ${\bm h}({\bm k})/ |{\bm h}({\bm k})|$ and is 
given by 
\begin{equation}
\nu  =  \iint_{BZ} \frac{d k_x d k_y}{2\pi}  
\left\{\partial_{k_x} a_y (k_x,k_y) -    
\partial_{k_y} a_x (k_x,k_y)\right\}. 
\label{eq:Chern}
\end{equation}
Equations (\ref{eq:pol}) and (\ref{eq:Chern}) lead to the 
relation~\cite{Dim,QiHuges,AliceaReview,Turner} 
\begin{equation}
\nu_{x} + \nu_{x}^{'} = \nu_y + \nu_y' = \nu \mod 2. 
\end{equation}

\begin{figure}
\includegraphics[height=7cm,width=7cm]{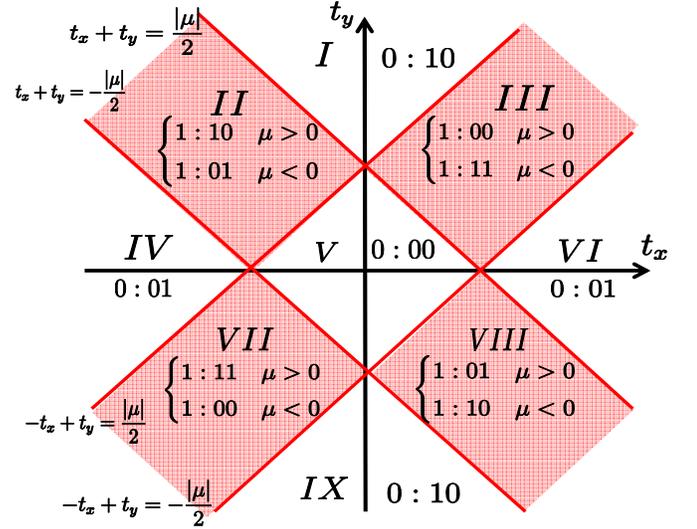}
\caption{The topological phase diagram of a model in eq.~(\ref{eq:p+ip}) for chiral superconductors in 2d characterized by 
$\nu:\nu_{x}\nu_{y}$. The lines where the energy gap closes divide the ($t_{x}$,$t_{y}$)-plane 
into nine domains. The system is the strong topological superconductors ($\nu=1$)
in the domains $I\!I$, $I\!I\!I$, $V\!I\!I$ and $V\!I\!I\!I$, while it is weak topological superconductor in 
the domains $I$, $I\!V$, $V\!I$ and $I\!X$ with $\nu=0$ but some of $Z_2$ invariants being nonzero.
In the domain $V$, the system is the trivial strong coupling superconductor.}
\label{fig:phase}
\end{figure}

In summary, our model is characterized by $Z$ topological 
invariant $\nu$ and two $Z_2$ topological invariants $\nu_{x}$ and $\nu_{y}$.
This is the general result, and superconductors in class $D$ in 2d 
are characterized by $\nu:\nu_{x}\nu_{y}$.
From the above consideration, these topological invariants depend
on the hopping integrals $t_x$, $t_y$ and the chemical potential $\mu$,
while do not depend on the pairing amplitudes $d_x$, $d_y$ as long as 
they are finite. Therefore, we show in Fig.~\ref{fig:phase} the 
phase diagram of the present model in the plane of $(t_x, t_y)$ for fixed $\mu$. 
The lines where the energy gap closes divide the $(t_{x},t_{y})$-plane into
nine domains. Electronic states are characterized by topological invariants 
in each domain. 
When $\frac{| \mu | }{2}$ is larger than $|t_x| + |t_y|$, i.e., domain $V$, the pairing state
is topologically trivial since it corresponds to the strong coupling limit.
In the domain $I\hspace{-.75mm}I$, $I\hspace{-.75mm}I\hspace{-.75mm}I$, 
$V\hspace{-.75mm}I\hspace{-.75mm}I$ and $V\hspace{-.75mm}I\hspace{-.75mm}I\hspace{-.75mm}I$, 
electronic states have both the 
$Z$ and $Z_2$ topological invariants. Note that the sign of the chemical potential 
$\mu$ matters for the $Z_2$ invariants.   
In the domain $I$, $I\hspace{-.75mm}V$, $V\hspace{-.75mm}I$ and $I\hspace{-.75mm}X$, 
the anisotropy between $t_x$ and $t_y$ is large
and hence the system behaves basically as the weakly coupled chains of 1d Kitaev models,
and the system has only 1d $Z_2$ topological invariants 
but $Z$ topological invariant $\nu=0$, 
so they are weak topological states.

\begin{figure}
\begin{flushleft}
\includegraphics[height=8cm,width=4cm,angle=-90]{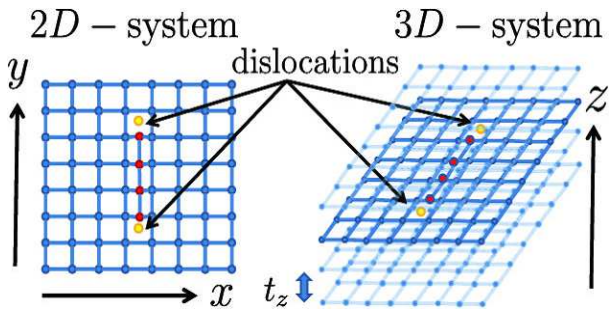}
\end{flushleft}
\caption{The edge dislocations (indicated by yellow points) in two dimensional system 
and three dimensional system.  
3d system is made by stacking 2d systems 
and adding a hopping integral along the z-direction with the edge dislocation
isolated on a layer.}
\label{fig:lattice}
\end{figure}

Now we turn to the consequences of the topological invariants. 
The $Z_2$ invariant $\nu_{x}$ ($\nu_{y}$)
ensures that the  propagating Majorana fermion channels 
appear at the edge along $x$ direction ($y$ direction).
They have zero-energy states at $k_{x}=\pi$ ($k_{y}=\pi$).
Also the edge dislocation offers the stage for the zero energy Majorana 
bound state when the following equation is satisfied (see Fig.~\ref{fig:lattice}).
\begin{equation}
\mathbf{B}\cdot\mathbf{G} = 1 \enskip \mod 2 \label{condi}. 
\end{equation}
In this equation, we define 
${\mathbf G}=\frac{1}{2\pi}(\nu_{x}{\mathbf b}_{x}+\nu_{y}{\mathbf b}_{y})$.
${\mathbf b}_{x}$ and ${\mathbf b}_{y}$ are reciprocal lattice vectors of 
$x$- and $y$-direction, respectively, and 
${\mathbf B}$ is the Burgers vector characterizing the dislocation.
We numerically calculate these zero-energy states as follows.
To introduce a periodic boundary condition, 
we introduce two edge dislocations with the Burgers vector 
${\mathbf B}=\pm {\mathbf e}_{x}$.
We represent two edge dislocations by adding lattice sites between them.
Edge dislocations are separated by a half system size. 
Calculations were done on a 40$\times$40 unit cell
system with a periodic boundary condition along $x$- and $y$-directions. 
The parameters are 
$t_{x} = 0.5$, $t_{y} = 0.5$, $d_{x} = 0.6$, $d_{y} = 0.6$, $\mu = -0.2$.
In these parameters, the topological invariants are $1:11$.
In this case: ${\mathbf G}=\frac{1}{2\pi}({\mathbf b}_{x}+{\mathbf b}_{y})$ 
and ${\mathbf B}=\pm {\mathbf e}_{x}$, 
${\mathbf B}\cdot{\mathbf G} = 1$ is satisfied, so zero-energy 
states appear at edge dislocations.

\begin{figure}
\begin{flushleft}
\includegraphics[height=4cm,width=9cm]{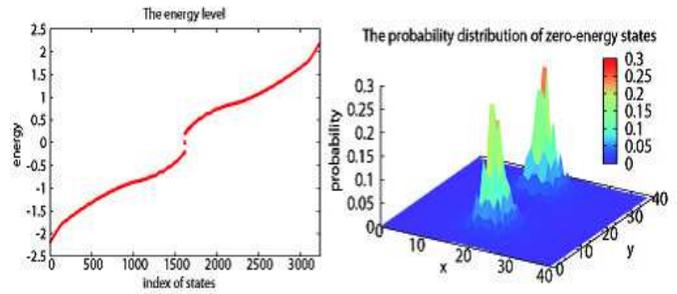}
\end{flushleft}
\caption{The energy levels and the probability distribution of 
zero-energy states at each 
site in the presence of edge dislocations: The left figure indicates 
the energy levels. The right figure indicates the probability distribution for
the zero-energy states. The zero energy Majorana bound 
state appear at each of the edge dislocations.}
\label{fig:Emap}
\end{figure}

The results of our numerical calculations are shown in Fig.~\ref{fig:Emap}.
The left figure of Fig.~\ref{fig:Emap} indicates the energy levels of our 
model in the presence of edge dislocations.
It is clear that zero-energy states exist, which are 2-fold 
degenerate because the 2 edge dislocations are present in the system. 
The right figure of Fig.~\ref{fig:Emap} indicates the probability 
distribution of the zero-energy 
states. Zero-energy states are localized at each edge dislocation.

In the presence of edge dislocations with the Burgers vector 
${\mathbf B}=\pm {\mathbf e}_{x}$,
zero-energy states appear in the domain 
$I\hspace{-.75mm}$,$I\hspace{-.75mm}I$,$V\hspace{-.75mm}I\hspace{-.75mm}I$ 
and $I\hspace{-.75mm}X$ 
in Fig.~\ref{fig:phase} in the case of $\mu>0$ 
and zero-energy states appear in the domain $I$,$I\hspace{-.75mm}I\hspace{-.75mm}I$,$V\hspace{-.75mm}I\hspace{-.75mm}I\hspace{-.75mm}I$ and $I\hspace{-.75mm}X$ 
in Fig.~\ref{fig:phase} in the case of $\mu<0$.
In particular, zero-energy states in the domain $I$ and $I\hspace{-.75mm}X$ 
can be intuitively interpreted as follows. 
The weak topological superconductors in these domains are adiabatically 
connected to a stack of the 
Kitaev models~\cite{Kitaev} for a 1d topological superconductor along $y$-direction.
In the presence of edge dislocations, the edges of 1d topological superconductor
appear at edge dislocations as shown in Fig.~\ref{fig:lattice}, so zero-energy states 
appear there.  In general, the existence of zero-energy states is proved by the 
same method as in ref.~\cite{Ran}.

\begin{figure}
\begin{flushleft}
\includegraphics[height=5.5cm,width=9cm]{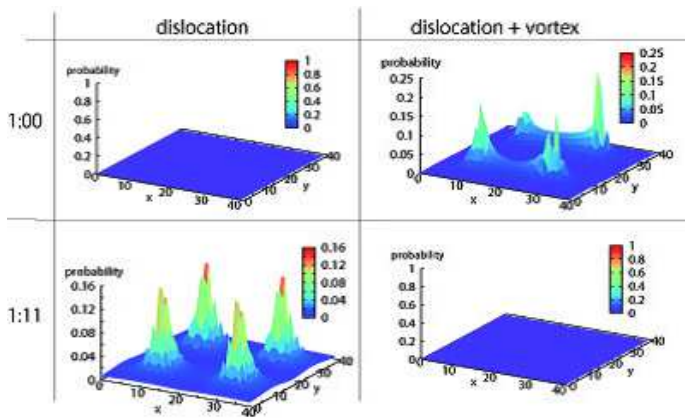}
\end{flushleft}
\caption{The interaction between zero-energy states localized at edge dislocations 
and zero-energy states localized at vortex cores: 
This figure shows the probability distribution of zero-energy 
states at each lattice sites in each case.
The interaction between zero-energy states at edge 
dislocations and zero-energy states at vortex cores
eliminates zero energy states when dislocations and 
vortex cores exist at the same positions.
}
\label{fig:vor+dis}
\end{figure}

Next we consider the interference of the $Z$ and $Z_2$ topological invariants, which is realized in the present model. 
When $Z$ is nonzero, the zero energy Majorana bound state is realized at the
core of the vortex~\cite{Read,Ivanov}. It is expected that, if the dislocations are
in the crystal, they act as the pinning centers of the vortex, and hence
there are two reasons for the existence of the Majorana bound states
when $Z_2$ invariant is 1.
This situation occurs in the domains $I\hspace{-.75mm}I$, $I\hspace{-.75mm}I\hspace{-.75mm}I$, $V\hspace{-.75mm}I\hspace{-.75mm}I$, and $V\hspace{-.75mm}I\hspace{-.75mm}I\hspace{-.75mm}I$ 
in Fig.~1,
and the interference of these two mechanisms is an issue.
Figure~\ref{fig:vor+dis} summarizes the calculated results,
in which there are dislocations and vortex cores at same positions.
The probability distributions are plotted for the zero energy
states if any.
To introduce a periodic boundary condition, 
we introduce two vortices with the winding number 1 and two 
vortices with the winding number -1.
We consider the two cases of topological invariants $1:00$ (upper panels) and 
$1:11$ (lower panels). The parameters are 
$t_{x} = 0.5$, $t_{y} = 0.5$, $d_{x} = 0.6$, $d_{y} = 0.6$, $\mu = 0.2$
for the former, while they are 
$t_{x} = 0.5$, $t_{y} = 0.5$, $d_{x} = 0.6$, $d_{y} = 0.6$, $\mu = -0.2$ for the latter.
Calculations were done on a 40$\times$40 unit cell system with a periodic 
boundary condition along $x$ and $y$-directions. 
There are dislocations and vortex cores at same positions in the 
right panels while only dislocations are there in the left panels.
They are separated from each other by a half system size.
In the case of the topological invariants $1:00$,
zero-energy states do not appear at edge dislocations.
Zero-energy states appear when the dislocations and vortices exist 
at the same time because of the $Z$ invariant $\nu=1$ and vortices. 
We have also confirmed that the zero-energy states appear with only the vortices.
In the case of the topological invariants $1:11$,
zero-energy states appear when dislocations exist but vortices do not exist.
The interaction between zero-energy states at edge dislocations and at vortex cores
eliminates zero energy states when they coexist at the same position. 
This can be naturally understood that the two Majorana bound states 
due to $Z$ and $Z_2$ invariants interact with each other as approaching to 
each other, and lift the degeneracy to have finite energies.

We generalize this model into the model in 3d.
The 3d model is constructed as a stack of 2d models 
with a hopping integral $t_z$ along the z-direction as shown in Fig.2.
$t_{z}$ causes the finite region where the 
energy gap closes in the $(t_{x},t_{y})$-plane.
The energy gap closes in the domain which satisfies 
$\vert t_{x}+t_{y}-\frac{\mu}{2} \vert < \vert t_{z} \vert$,
$\vert -t_{x}+t_{y}-\frac{\mu}{2} \vert < \vert t_{z} \vert$,
$\vert t_{x}-t_{y}-\frac{\mu}{2} \vert < \vert t_{z} \vert$ or 
$\vert -t_{x}-t_{y}-\frac{\mu}{2} \vert < \vert t_{z} \vert$.
The domain where the energy gap opens is connected to 
the domain of $t_{z}=0$, so electronic states in the domain have
the same topological invariants as the 2d model, while the topology is trivial for class D in 3d.
We introduce the similar defect as the 2d system 
as indicated in Fig.~\ref{fig:lattice}.
These defects are not edge dislocations. One layer has 
the same defect as the 2d system and other layers 
do not have it. 
We ensure that zero-energy states appear at those defects 
in the same manner 
as the 2d system by the numerical calculations. 
The parameters are $t_{x} = 0.5$, $t_{y} = 0.5$, $t_{z} = 0.1$, $d_{x} = 0.6$, 
$d_{y} = 0.6$, $\mu = -0.5$. In this case, the electronic state is adiabatically 
connected to the state in the domain $III$ in the case of $\mu < 0$ in Fig.~\ref{fig:phase}.
If $t_{z}=0$, zero-energy states appear in these parameters because the state in the 
domain $III$ in the case of $\mu < 0$ in Fig.~\ref{fig:phase} has topological 
invariants $1:11$ and one layer has edge dislocations with the Burgers vector 
$\mathbf{B}= \pm \mathbf{e}_{x}$.
Calculations were done on a 10$\times$20$\times$10 unit cell system 
with a periodic boundary condition along $x$, $y$ and $z$-directions.
We found that these zero-energy states survive a finite-value z direction hopping integral $t_{z}$ 
unless the energy-gap closes. 
This edge dislocation is 0d object, and we need $D=2$ sphere to enclose 
this defect in 3d, and hence $\delta=3-2=1$. Therefore, the $Z_2$ invariant 
protects the existence of the Majorana zero energy bound state.
This is an example of the "weak-weak" topological superconductor,
where the topological invariant reduced by 2 dimensions is relevant.
This indicates that Majorana fermions can exist even in a 3d system. 

Now we discuss the relevance of the present results to the
real materials. There are several candidates for the chiral 
superconductors.  The 2d Rashba system in a semiconductor with the proximity to
the $s$-wave superconductor and the ferromagnet is a promising candidate, 
which shows the spinless $p+ip$ pairing~\cite{Fujimoto,Sau,Lutchyn,Alicea}. 
The electron density is considered to be small and concentrated near 
the $\Gamma$-point of the 1st Brillouine zone. Therefore, usually the continuum 
approximation is used to describe this system. In our phase diagram (Fig.~\ref{fig:phase}), 
this situation corresponds to the domain $VII$ with $t_x = t_y < 0$ with negative chemical potential,
which is characterized by $\nu=1$ and $\nu_x=\nu_y=0$. Therefore, we do expect the
Majorana bound state at the core of the vortex, while there is no Majorana
bound state at the edge dislocation. To realize more interesting situation where
both  $\nu$ and $\nu_{x,y}$ are nonzero, it is required to have the Fermi pocket near 
the zone boundary or zone corner. This is, in principle, possible once the 
sign of the hopping integral is reversed.    
Another interesting candidate is Sr$_2$RuO$_4$~\cite{SRO}, which is believed to
be a quasi-2d $p+ip$ superconductor with the non-zero $Z$ invariant.
In this system, there remains the spin degeneracy of the Fermi surface, 
and hence the Hamiltonian matrix is at least 4$\times$4 instead of 2$\times$2 
discussed in this paper. This means that the $Z_2$ invariants are zero
due to this degeneracy, although Z invariant can be nonzero. 

To summarize, we have studied a model of chiral superconductors
including both the Kitaev model in 1d and $p+ip$ superconductor in 2d
as limiting cases. This model shows a rich phase diagram (Fig.~1)
characterized by the $Z$ and $Z_2$ topological invariants, which
control the appearance of Majorana bound states at the
edge dislocations and vortex cores. 
This offers an explicit case where the topological periodic table 
can be successfully applied including the topological textures, and the
presence of the Majorana bound states are shown also numerically.

The authors acknowledge the fruitful discussion with Yukio Tanaka, and 
Sho Nakosai. 
This work is supported by Grant-in-Aid for Scientific Research
(Grants No. 17071007, No. 17071005, No. 19048008,
No. 19048015, No. 22103005,
No. 22340096, and No. 21244053) 
from the Ministry of Education, Culture,
Sports, Science and Technology of Japan, Strategic
International Cooperative Program (Joint Research Type)
from Japan Science and Technology Agency, and Funding
Program for World-Leading Innovative RD on Science and
Technology (FIRST Program).

\end{document}